\documentclass[a4paper,11pt]{article}
\usepackage[latin1]{inputenc}
\usepackage[T1]{fontenc}
\usepackage[english]{babel}
\usepackage{graphicx}
\usepackage{float}
\begin{document}
\title{A scientometric study of General Relativity and Quantum Cosmology from 2000 to 2012}
\author{Stéphane Fay\footnote{steph.fay@gmail.com}\\
Palais de la Découverte\\
Astronomy Department\\
Avenue Franklin Roosevelt\\
75008 Paris, France\\
and\\
Sébastien Gautrias\\
IT\\
St Maur des Fossés 94, France}
\maketitle
\begin{abstract}
$2015$ is the centennial of Einstein General Relativity. On this occasion, we examine the General Relativity and Quantum Cosmology (GRQC) field of research by analysing $38291$ papers uploaded on the electronic archives arXiv.org from $2000$ to $2012$. We establish a map of the countries contributing to GRQC in $2012$. We determine the main journals publishing GRQC papers and which countries publish in which journals. We find that more and more papers are written by groups (instead of single) of authors with more and more international collaborations. There are huge differences between countries. Hence Russia is the country where most of papers are written by single authors whereas Canada is one of the countries where the most of papers imply international collaborations. We also study authors mobility, determining how some groups of authors spread worldwide with time in different countries. The largest mobilities are between USA-UK and USA-Germany. Countries attracting the most of GRQC authors are Netherlands and Canada whereas those undergoing a brain drain are Italy and India. There are few mobility between Europe and Asia contrarily to mobility between USA and Asia.
\end{abstract}
\section{Introduction} \label{s0}
\begin{figure}[p]
\centering
\includegraphics[width=20cm,angle =90]{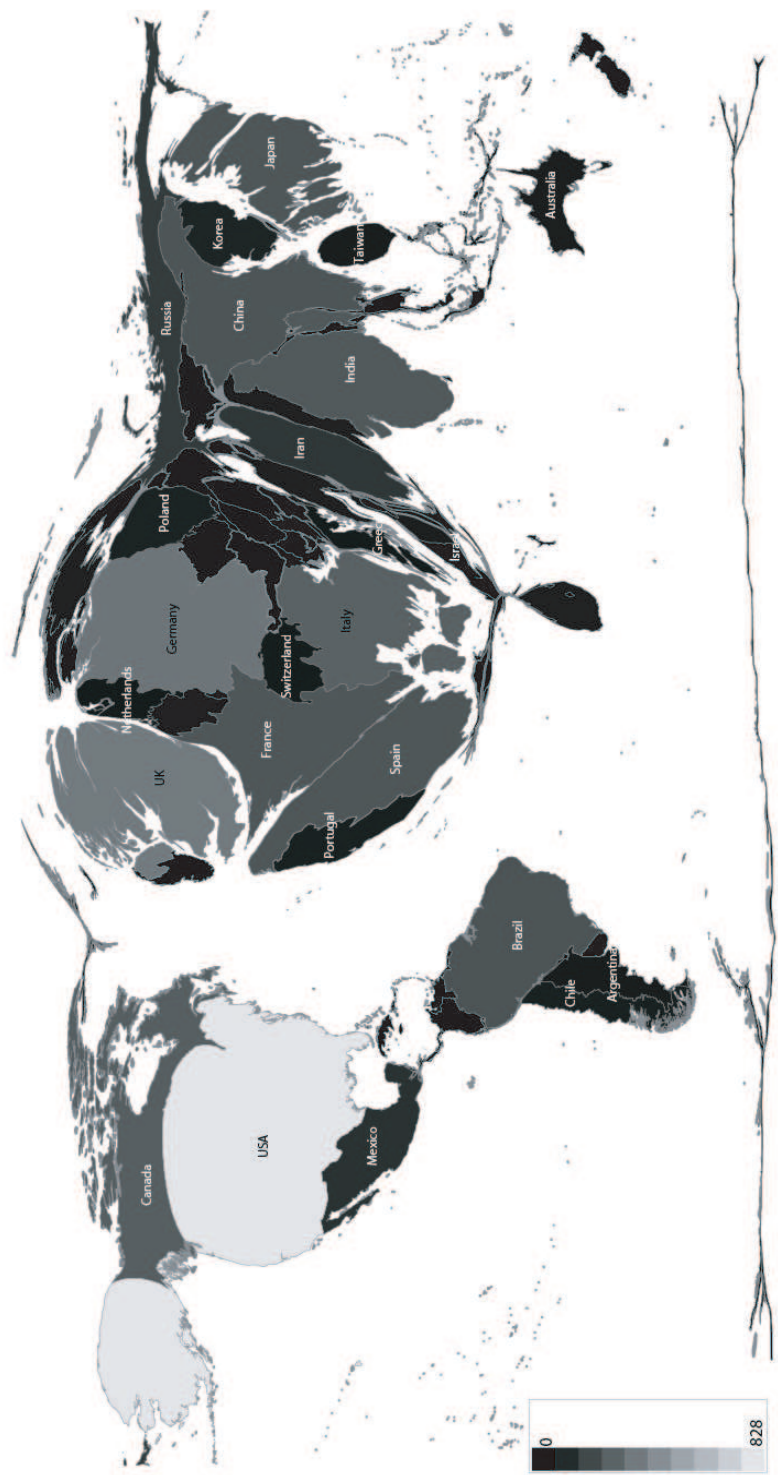}
\caption{\scriptsize{\label{fig0}Map representing the number of papers deposited on the gr-qc archive with respect to the country where their authors were affiliated in $2012$. Countries borders and colors are distorted (cartogram) by this number: the larger the number of papers, the larger the size of a country, the brighter its color. Only the names of countries whose affiliated authors have deposited more than $50$ papers on the gr-qc archive are indicated.}}
\end{figure}

$2015$ is the centennial of Einstein General Relativity\cite{Ein15}. To mark this centennial, this paper presents a scientometric study of General Relativity and Quantum Cosmology (GRQC) from $2000$ to $2012$. General Relativity is the gravitation theory that allows to describe Universe evolution. Quantum Mechanics\cite{Pla00} is the theory that describes the world of particles. Physicists hope to merge these two theories to get a Quantum Gravity theory that, when applied to the Universe, is called Quantum Cosmology. Quantum Cosmology should help us to better understand Universe, in particular when it is very tiny, near the Big Bang singularity. GRQC is thus an important field of research and we want to determine to which countries are affiliated the GRQC authors, in which journals they publish GRQC papers and how they interact the ones with the others.\\
To reach this goal, we use the "gr-qc" papers database of the electronic archive arXiv.org\cite{grqc}. This is a large collection of papers from $1992$ to today about GRQC. Let us note that more and more researchers use some electronic archives. Hence in Canada, $53\%$ of Canadian researchers in physics and astronomy use specific search engines for searching literature and $52\%$ of their papers are deposited in a free repository\cite{Can14}. For our study, we consider papers deposited on the gr-qc archive from years $2000$ to $2012$ (see appendix \ref{a0} for the data strategy retrieval and data summary). They consist in $38291$ papers written by $16055$ authors and published in $463$ journals. We do not consider papers before $2000$ because they are too few to make viable statistical analysis nor beyond $2012$ because a large number of papers beyond this date has not been yet referenced with a journal (more than $50\%$ at the time when we write these lines). For each of these $38291$ papers, we determine the journal that publishes it (if any), the names of its authors and the countries to which they are affiliated. For each author, this last one is defined as the country of his/her host institution as indicated in each of his/her papers (when several institutions are indicated in a paper, we consider that the first one is the host institution). Then we study these data to get a picture of GRQC research worldwide. Note that one difficulty about study of data coming from the internet is the influence of authors habits in using electronic archives to deposit their papers. For instance, if there are more papers deposited in the gr-qc archive for the year $2012$ than for the year $2000$, this is obviously not related to the development of GRQC during these $13$ years but to the increasing number of authors depositing their papers on the electronic archive. We keep this bias in mind all along this study.\\
To understand the importance of GRQC in the world today, we first establish the map \ref{fig0}. It shows the number of papers deposited on the gr-qc archive with respect to countries where authors were affiliated in $2012$. Hence, in $2012$, authors affiliated to USA are the largest contributors to the gr-qc archive with $828$ papers deposited on it. This map allows thus to compare, at a glance, the contribution of each country in the world to GRQC field of research that we are going to study all along this paper whose plan is the following.
In the second section, we analyze which journals publish GRQC papers. We show among other results that only $2$ of them publish around $40\%$ of all the GRQC papers. In a third section, we consider authors. One result is that although much papers are written by single author, more and more papers are written by groups of authors over time with more and more international collaborations. This trend is already observed in other fields of research like biology\cite{Car14}. In the fourth and fifth sections, we examine collaborations between authors depending on the countries to which they are affiliated and their mobility. By far, the largest collaborations and mobilities occur between UK/USA and Germany/USA, whereas we observe a brain drain from Italy. We conclude in the last section.
\section{Journals} \label{s11}
\begin{figure}[H]
\centering
\includegraphics[width=8cm]{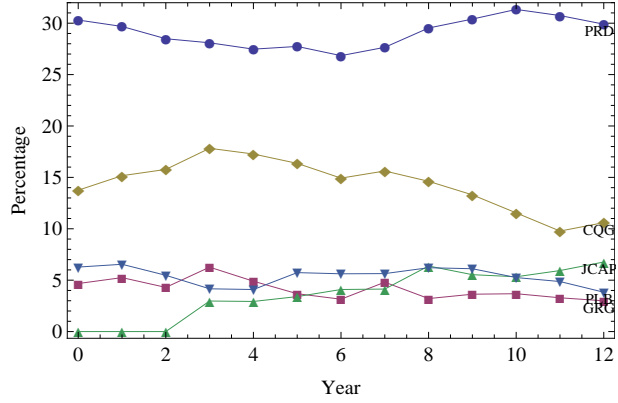}
\caption{\scriptsize{\label{fig6} Five journals publish the most of GRQC papers: Physical Review D (PRD), Classical and Quantum Gravity (CQG), Journal of Cosmology and Astroparticles (JCAP), General Relativity and Gravitation (GRG), Physics Letters B (PLB). The graph shows, for the years $2000$ to $2012$, the percentages of published GRQC papers deposited on the gr-qc archive that each of these journals have published. It is stable for PRD, decreasing for CQG and increasing for JCAP (first edited in $2003$).}}
\end{figure}

From the analysis of the gr-qc archive, we found that $463$ journals have published papers about GRQC between $2000$ to $2012$. However, only $16$ of them published around $72\%$ of all these GRQC papers (see appendix \ref{a1} for a list of these journals and the percentages of papers they published in $2000$, $2004$, $2008$ and $2012$). Each of the others journals published less than $1\%$ of all the GRQC papers deposited on the gr-qc archive each year. Two journals are of prime importance since they published around $40\%$ of GRQC papers: "Physical Review D" (PRD) and "Classical and Quantum Gravity"(CQG). We remark that the number of GRQC papers published by CQG is decreasing. It could be overcame in some few years by "Journal of Cosmology and Astroparticles" (JCAP) as shown on figure \ref{fig6}, a more recent journal (from the same editor IOP Science). Furthermore, among countries whose affiliated authors have published more than $100$ papers in $2012$, some countries already published more papers in JCAP than in CQG. Note finally that the journal in which all the countries publish the most of papers is PRD as shown on table \ref{tab1}.
\section{Authors} \label{s12}
\begin{figure}[H]
\centering
\includegraphics[width=6cm]{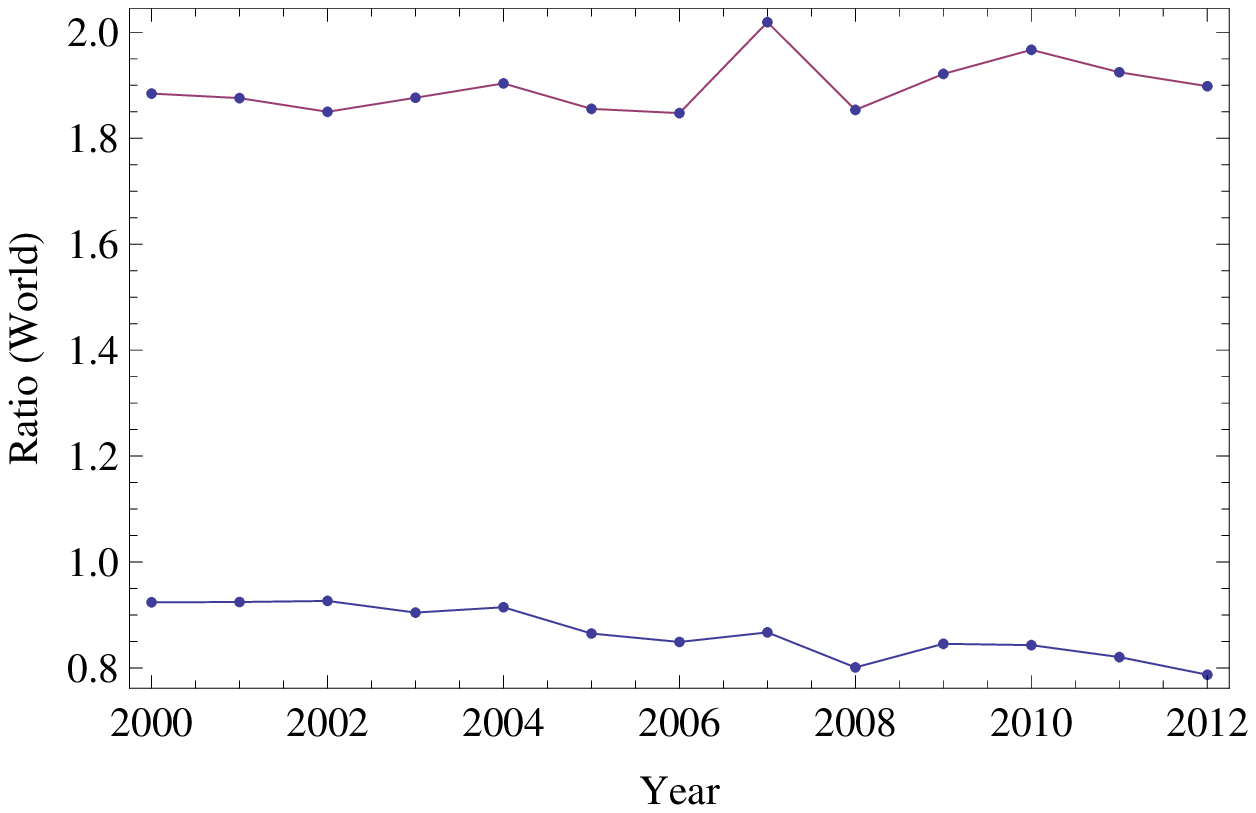}
\includegraphics[width=6cm]{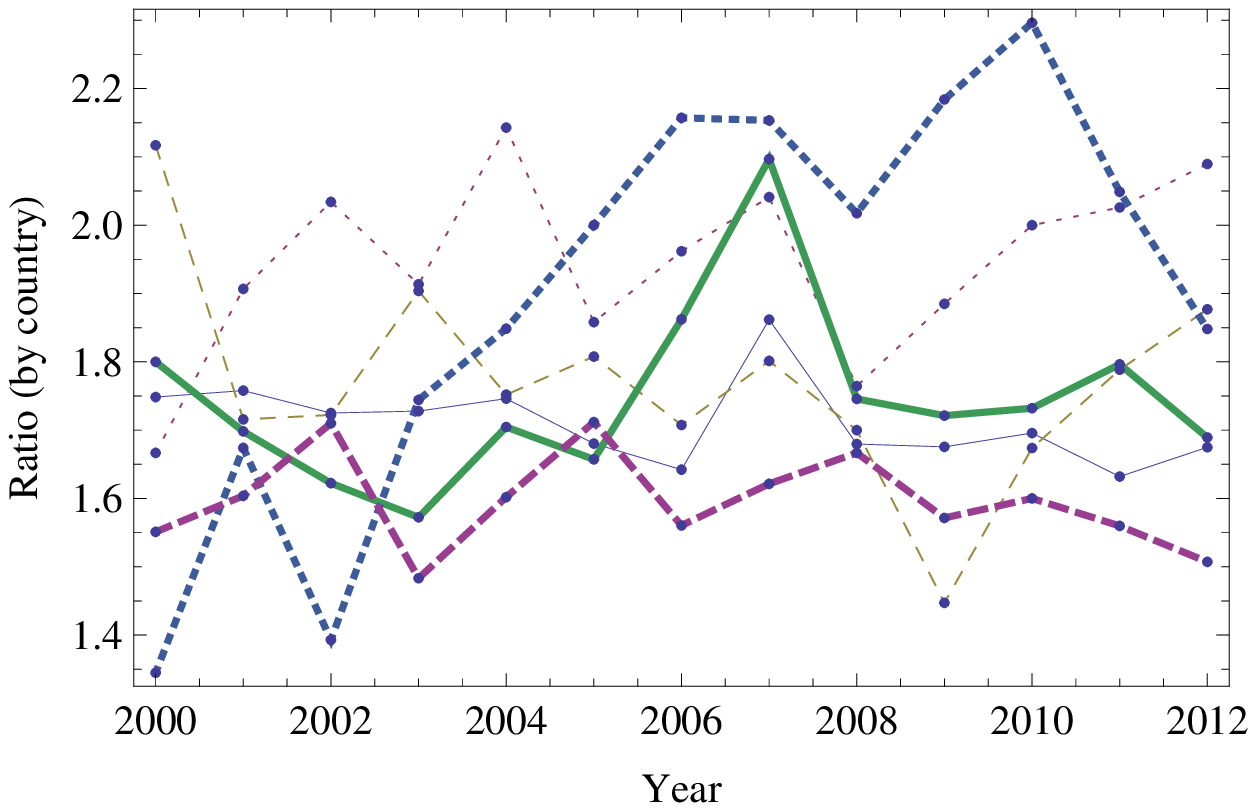}
\includegraphics[width=8cm]{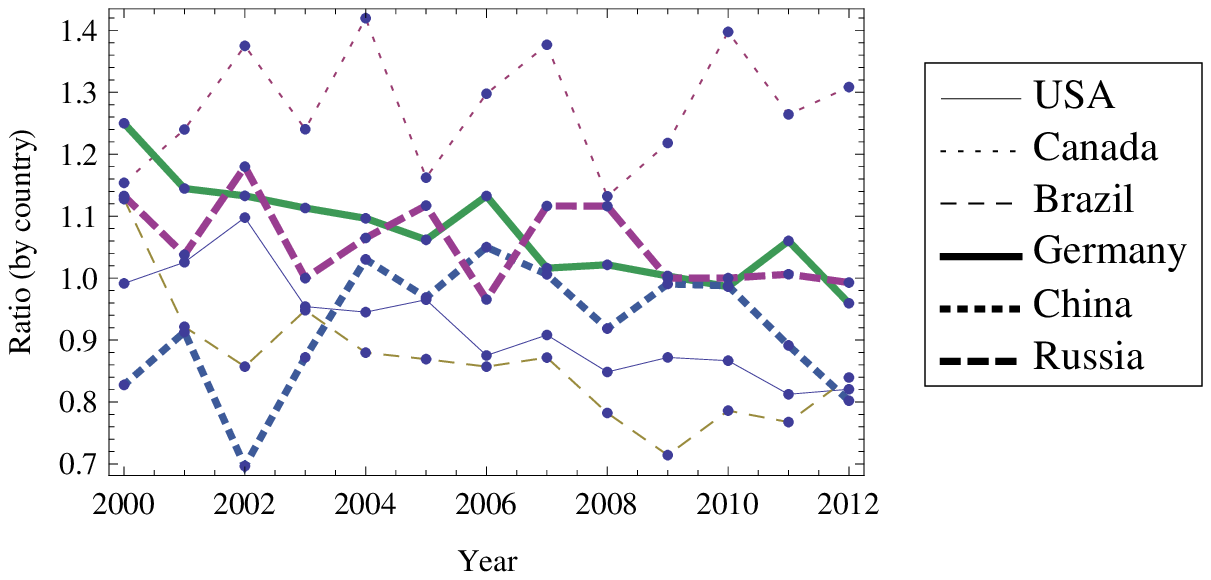}
\caption{\scriptsize{\label{fig4}The first graph shows the average number of papers written by each author (upper curve) and density of papers by author (lower curve) for all the world from $2000$ to $2012$. The second and third graphs show each of these two quantities for several countries.}}
\end{figure}

\begin{figure}[H]
\centering
\includegraphics[width=6cm]{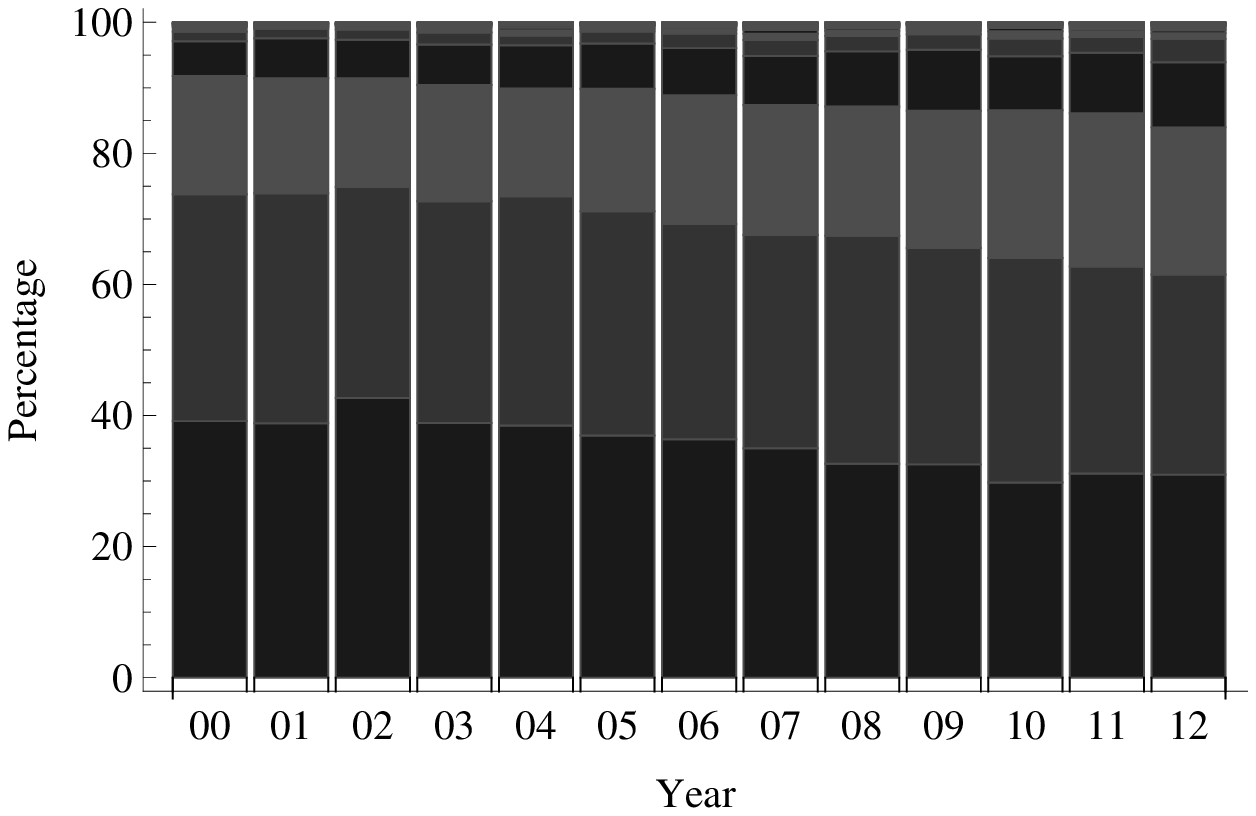}
\includegraphics[width=6cm]{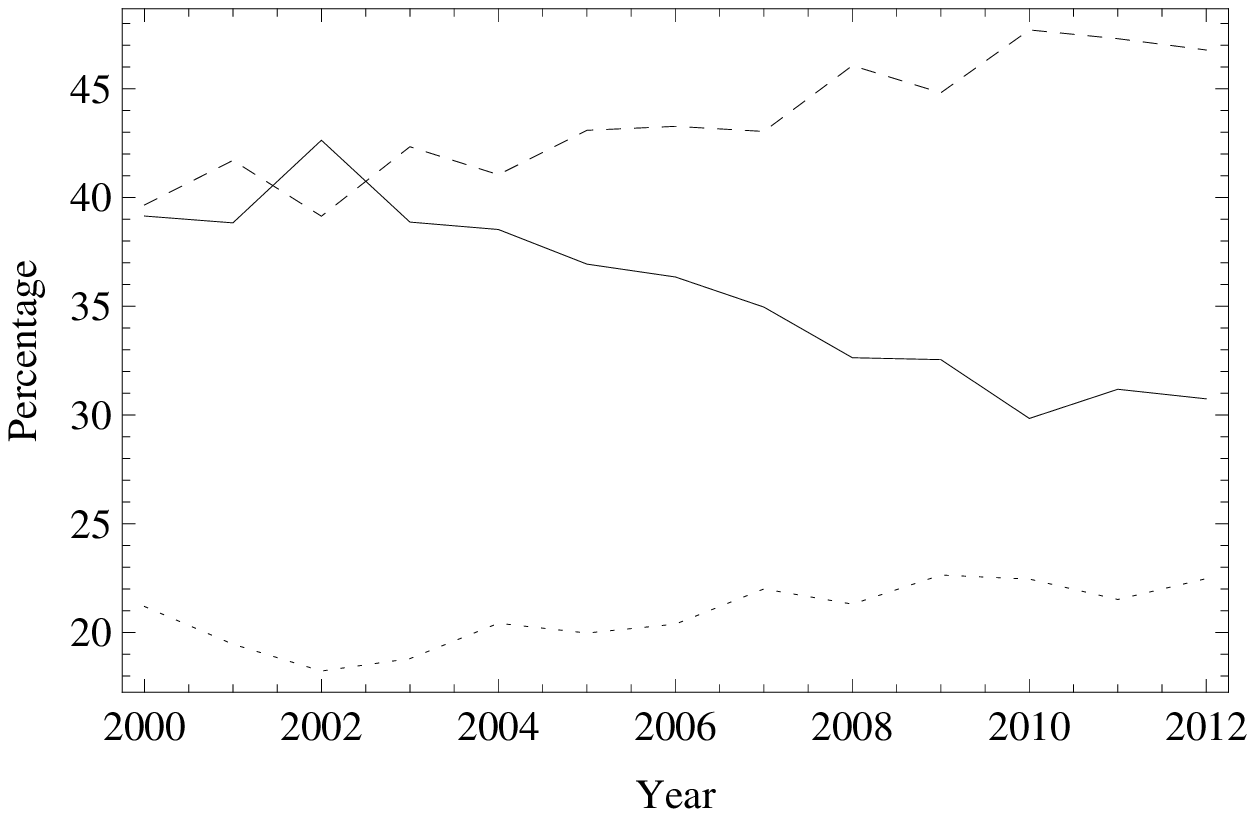}
\caption{\scriptsize{\label{fig2}First graph: percentages of papers written by $1$ (the lowest part of each column) to $4$ (the fourth part of each column) and more authors from $2000$ to $2012$. Second graph: percentages of papers written by $1$ author (thin), several authors affiliated to the same country (dashed) or several authors affiliated to different countries (dotted) from $2000$ to $2012$.}}
\end{figure}

We found $16055$ authors that wrote papers about GRQC and deposited them on the gr-qc archive from $2000$ to $2012$. A map very similar to that of figure \ref{fig0} can be obtained to show the geographical distribution of these authors in the world for the year $2012$.\\
For each year from $2000$ to $2012$, we calculate the total number of papers deposited over the total number of authors of these papers. We call this quantity the density of papers by author. It is shown on the first graph (lower curve) of figure \ref{fig4}. For the world, it varies between $0.77$ and $0.92$. It is different for each country as shown on the third graph of figure \ref{fig4}. Hence, the country whose affiliated authors have the largest density from $2001$ to $2012$ is Canada. It is thus the country where increasing the number of affiliated GRQC authors will increase the most the number of papers published by this country.\\
\begin{table}[H]
\begin{center}
\begin{tabular}{|r|c|c|c|c|}
\hline
Country & First journal & Second journal & Third journal& Total\\
\hline
USA & PRD ($41\%$) & CQG ($11\%$) & PRL ($7.4\%$) & $59\%$\\
Canada & PRD ($33\%$) & CQG ($12\%$) & PRL ($7.9\%$) & $53\%$\\ 
UK & PRD ($41\%$) & CQG ($14\%$) & JCAP ($13\%$) & $68\%$\\
France & PRD ($38\%$) & CQG ($18\%$) & JCAP ($13\%$) & $69\%$\\
Germany & PRD ($26\%$) & CQG ($14\%$) & MNRAS ($11\%$) & $51\%$\\
Japan & PRD ($40\%$) & JCAP ($16\%$) & CQG ($7.8\%$) & $64\%$\\
Spain & PRD ($34\%$) & JCAP ($12\%$) & CQG ($12\%$) & $58\%$\\
Brazil & PRD ($32\%$) & JCAP ($7.9\%$) & CQG ($7.0\%$) & $47\%$\\
Russia & PRD ($22\%$) & CQG ($11\%$) & Grav. $\&$ Cos. ($11\%$) & $44\%$\\
Italy & PRD ($16\%$) & JCAP ($9.6\%$) & JHEP ($8.0\%$) & $34\%$\\
China & PRD ($31\%$) & JHEP ($9.8\%$) & Eur. Phys. J ($8.4\%$) & $49\%$\\
India & PRD ($32\%$) & IJMPD ($7.8\%$) & Phys. letters B ($6.8\%$) & $47\%$\\
Iran & PRD ($21\%$) & Astr. and Spa. Sc. ($13\%$) & JCAP ($8.8\%$) & $43\%$\\
Mexico & PRD ($33\%$) & AIP Conf. Proc. ($9.8\%$) & JCAP ($9.8\%$) & $53\%$\\
\hline
\end{tabular}
\caption{\scriptsize{For each country whose affiliated authors deposited more than $100$ papers on the gr-qc archive in $2012$, this table shows the $3$ main journals in which they published these GRQC papers with the corresponding percentage with respect to the total number of papers they deposited this year.}}
\label{tab1}
\end{center}
\end{table}
Another interesting quantity is the average number of papers written by or to which collaborate each author. It differs from the density of papers by author: for instance if a group of $10$ authors writes $10$ papers, the density for this group is then $10/10=1$. However, each author can collaborate to several papers and the average number of papers written by or to which collaborated each author is thus larger. This is this last quantity that is shown for the world on the first graph (upper curve) of figure \ref{fig4}. It varies between $1.8$ and $2$ but is different for each country as shown on the second graph of figure \ref{fig4}. Hence, from $2004$ to $2011$, China is one of the country where each author contributes to the most of papers each year.\\
From a more general viewpoint, the number of authors from $2000$ to $2012$ increases faster than the number of papers (see appendix \ref{a0}). This is due to the fact that the number of authors collaborating to write a paper is increasing with time as shown on the first graph of figure \ref{fig2}. Whereas in $2000$, $39\%$, $34\%$, $18\%$ and $5\%$ of papers were written respectively by $1$ to $4$ authors, in $2012$ these proportions were respectively $30\%$, $30\%$, $22\%$ and $10\%$.\\
We also analysed the number of papers written by $1$ author, by several authors affiliated to the same country or by several authors affiliated to different countries (i.e. international collaboration) from $2000$ to $2012$. Results are shown on the second graph of figure \ref{fig2}. Whereas the first quantity decreases, the two last ones increase slowly during these $13$ years. The increasing need of networking for a successful career in science is likely to be one of the source of this tendency\cite{Pow12}. However, the situation is different for each country as illustrated by table \ref{tab3} in appendix \ref{a2} for the year $2012$ and for countries whose authors have deposited more than $50$ papers this year. Hence, Russia is the country whose affiliated authors write the most of papers alone: $50\%$ in $2000$, $46\%$ in $2006$ and $38\%$ in $2012$. China is the country whose affiliated authors write the most of papers with several authors affiliated to the same country (i.e. China): $64\%$ in $2003$, $56\%$ in $2006$ and $65\%$ in $2012$. China is thus among the countries whose affiliated authors have few collaborations with authors affiliated to other countries but the number of its international collaborations is increasing with time ($11\%$ in $2003$, $19\%$ in $2006$ and $22\%$ in $2012$). The authors that collaborate the more with authors affiliated to other countries than theirs are generally affiliated to European countries (for instance, Spain with $47\%$ of its papers in $2000$, $58\%$ in $2006$ and $57\%$ in $2012$) and Canada ($51\%$ of its papers in $2000$, $44\%$ in $2006$ and $50\%$ in $2012$). For comparison, in $2010$ a report from the European commission\cite{Eur10} indicated that the level of collaboration for all the European countries, for all the fields of research, with academic researchers from other countries was $47.8\%$.
\\
Surprisingly, there is not so much papers written in the USA that imply authors affiliated to other countries since only $33\%$ in $2000$, $32\%$ in $2006$ and $32\%$ in $2012$ of papers written by authors affiliated to USA implied international collaborations (by comparison, in $2012$, it was $48\%$ for UK or $50\%$ for Greece). It was already noticed in \cite{Eur10}, for all the fields or research, that the number of researchers not involved in formal collaboration with other countries was highest in the USA than for all the European countries. It could be due to the fact that the number of authors and institutions in the USA is so large that the necessity to collaborate with authors affiliated to other countries is less strong than anywhere else. Despite this, USA is a major partner of most countries in the world, as shown in the two next sections. In some way, if USA is attractive for the rest of the world, the opposite is not necessarily true.
\section{International collaborations} \label{s13}
\begin{figure}[p]
\centering
\includegraphics[width=16cm]{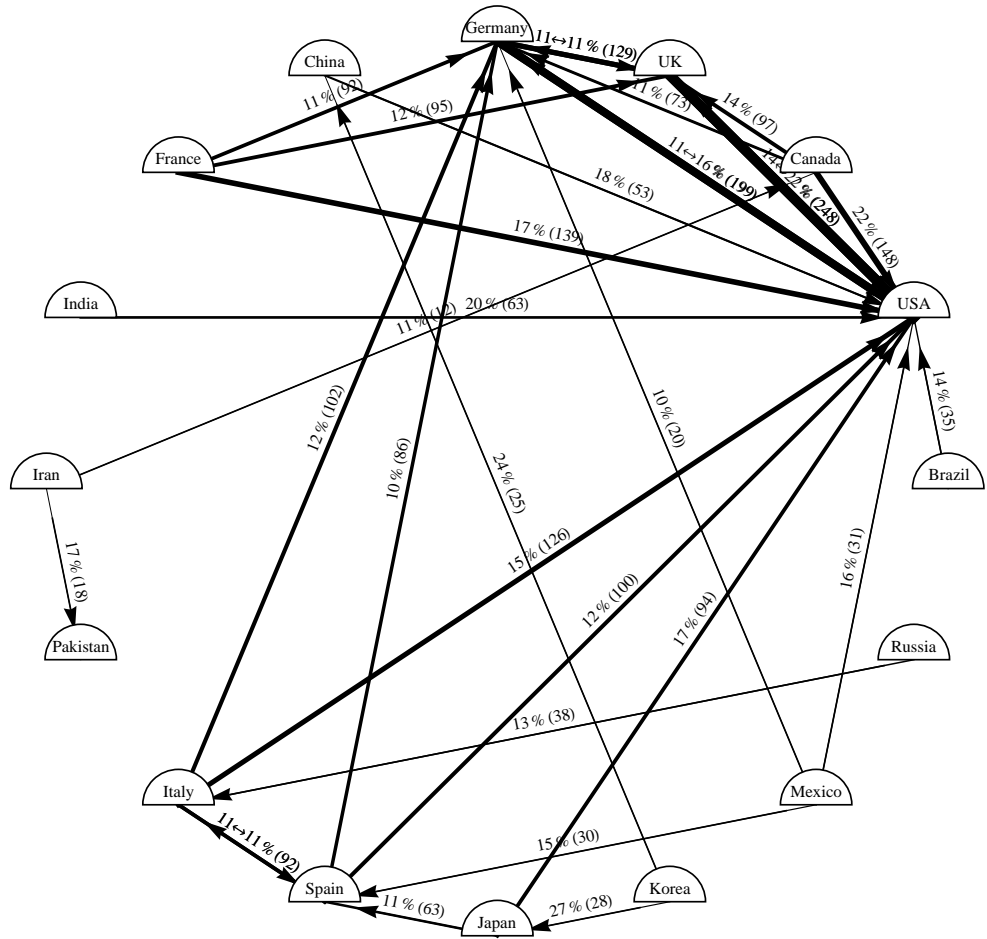}
\caption{\scriptsize{\label{fig7}On this graph, we show the countries whose affiliated authors have deposited more than $350$ papers on the gr-qc archive from $2008$ to $2012$ and the countries with which they wrote more than $10\%$ of their papers implying international collaborations. An arrow from a country "A" to a country "B" indicates that authors affiliated to country $A$ wrote more than $10\%$ of their papers implying international collaborations with authors affiliated to country $B$. The exact percentage and number of papers are then indicated above the arrow. The larger the line, the larger the number of papers on which authors affiliated to $A$ and $B$ have collaborated. For instance, the simple arrow going from Italy to USA indicates that USA is an important partner for Italy but that Italy is a marginal partner for USA, authors affiliated to these two countries having collaborated on $126$ papers during the years $2008-2012$, representing $15\%$ of Italy's international collaborations. The double arrow between Italy and Spain indicates that these two countries are important partners the one for the other, their affiliated authors having collaborated on $92$ papers that represents for both countries $11\%$ of their GRQC papers deposited on the gr-qc archive and implying international collaborations}}
\end{figure}

We analysed the international collaborations between authors affiliated to different countries during three periods, i.e. $\left[2000, 2004\right]$, $\left[2004, 2008\right]$ and $\left[2008, 2012\right]$. Results for each of these periods are very similar. We thus only present here the results got for the last period for the countries whose affiliated authors have deposited more than $350$ papers on the gr-qc archive during this $5$ years period. For these countries, we establish what are their main partners defined as the countries with which their affiliated authors wrote more than $10\%$ of their papers implying international collaborations. These results are presented on figure \ref{fig7}.
Its most striking feature is that authors affiliated to the USA collaborate nearly with authors affiliated to all the major countries producing papers in GRQC. Remembering section \ref{s12} where is was indicated that most of the papers written by authors affiliated to the USA did not imply international collaborations, this could sound like a paradox. It can be explained by the large number of authors affiliated to the USA and papers they write, with respect to the rest of the world. Even, if only $1/3$ of GRQC papers written by authors affiliated to the USA implies international collaborations (see appendix \ref{a2}), this number is sufficiently large so that USA be the major partner of most countries in the world.\\
We identified two others countries with which authors in the rest of the world collaborate the most after the USA, i.e. Germany and UK. The largest flows of papers implying international collaborations are between USA, Canada, European countries (UK, Germany, France, Italy, Spain) and Japan with major flows between USA-UK and USA-Germany. Between $2008$ to $2012$, authors affiliated to Iran\cite{Asn08} and Korea became important contributors to the gr-qc archive, each of them depositing on it more than $350$ papers during this $5$ years period. Authors affiliated to Iran mainly collaborate with authors affiliated to Pakistan and Canada whereas authors affiliated to Korea mainly collaborate with authors affiliated to China and Japan. In a general way, apart Iran and Korea, among the countries whose affiliated authors have deposited more than $350$ papers on the gr-qc archive from $2008$ to $2012$, the ones having the less of international collaborations during this period are Brazil, Russia, China and India (India had more international collaborations between $2000$ and $2008$ than from $2008$ to $2012$).
\section{Authors mobility} \label{s14}
\begin{figure}[h]
\centering
\includegraphics[width=12cm]{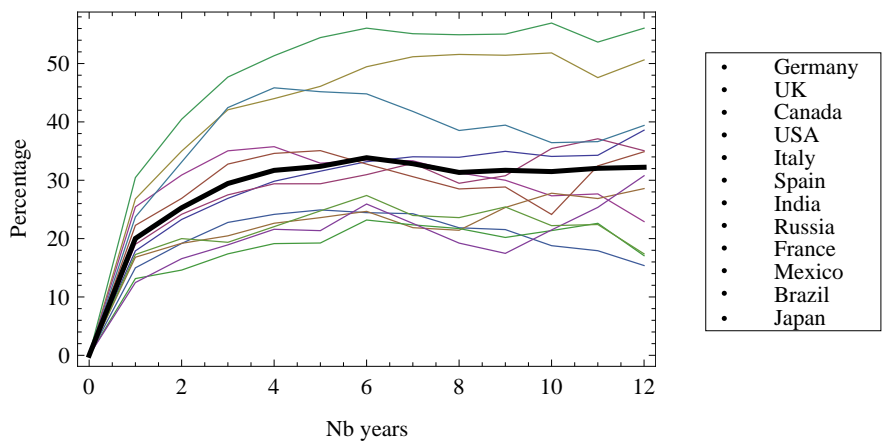}
\caption{\scriptsize{\label{fig8}Be a group of authors initially affiliated to a country, this figure shows the percentage of these authors still depositing papers on the gr-qc archive after $1$ to $12$ years but affiliated outside their initial country. We plotted here the curves for the $12$ countries having more than $50$ affiliated authors depositing papers on the gr-qc archive from $2000$ to $2012$. The thick curve gives the average of all the curves. Order of the countries in the column corresponds to order of the ends of each curve on the figure.}}
\end{figure}

We first study how a group of authors affiliated to the same country during the same year spreads worldwide with time. To this aim, we consider the $12$ countries having more than $50$ affiliated authors depositing papers on the gr-qc archive from $2000$ to $2012$. The figure \ref{fig8} shows, for groups of authors initially affiliated to the same country, the percentage of these authors still depositing papers on the gr-qc archive after $1$ to $12$ years but being affiliated in a country different from their initial country\footnote{Let us explain how we proceed to get this result by taking an example, i.e. USA. We first consider the initial group of authors having deposited papers on the gr-qc archive in $2000$ and being affiliated to the USA. We calculate for each year, between $2000$ and $2012$, the ratio of the number of authors of this group having deposited papers on the archive and being affiliated outside the USA over the number of authors of this group having deposited papers on the archive (being or not affiliated to the USA). We thus get a series of $13$ numbers representing the spread of the initial group during $12$ years. We repeat the same calculations for the initial group of authors affiliated to the USA and having deposited papers on the gr-qc archive in $2001$, thus getting a second series of $12$ numbers and so on, until considering a last group in $2011$. We then sum all these series so as to get a mean indicating the spread outside the USA of an initial group of authors affiliated to this country after $1$ to $12$ years. This gives the curves on figure \ref{fig8} obtained for $12$ countries}.
The thick curve represents the average for the $12$ countries. A possible interpretation of its shape is the following. Initially (when Nb years $=0$ on figure \ref{fig8}), all the authors (that are PhD, postdocs and permanent researchers) are affiliated to their initial countries and the percentage of authors affiliated outside these initial countries is thus $0$. The next years, mainly some PhD and postdocs leave their initial countries and this percentage increases. It reaches a maximum after $6$ years. Then, it slightly decreases before stabilising since authors affiliated outside and inside their initial countries get more and more permanent positions, their numbers thus tending to a constant. After an average of $8$ years, around $30\%$ of authors still depositing papers on the gr-qc archive have found a permanent position outside the initial country where they were affiliated. For comparison, note that in a recent survey $44.9\%$ of canadian postdoc (for all fields of research) hoped to get a stable full-time employment after $1$ to $3$ years as postdoc\cite{mitacs13}. If this interpretation is correct, these are clearly authors initially affiliated to Germany and UK that will most often get a permanent position outside these countries (more than $40\%$). The USA is close to the average: authors that have worked in the USA, and still deposit papers on the gr-qc archive $12$ years later, are around $1/3$ to be affiliated outside of the USA after $6$ years. This indicates that USA retain better its GRQC authors than UK or Germany (note that it was already shown in \cite{Eur10} that USA retain better its graduate (for all fields of science) than Europe). Last, we remark that authors initially affiliated to Japan, Brazil or Mexico will most often stay affiliated to these countries (at least $80\%$) that underlines a lack of mobility of these groups of authors with respect to other countries.

\begin{figure}[H]
\centering
\includegraphics[width=8cm]{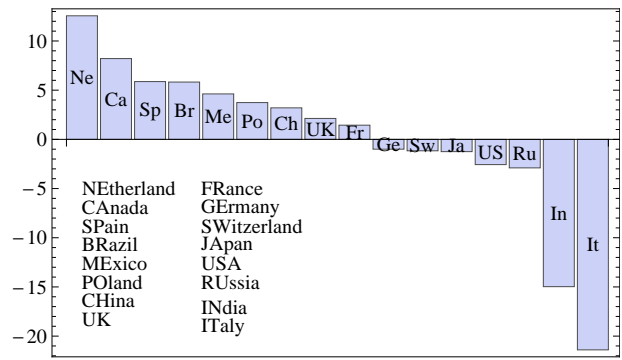}
\caption{\scriptsize{\label{fig9} Ratio (times $100$) of the difference between the number of authors coming in and out over the average of authors coming in and out for countries having more than $100$ authors coming in and out of them from $2000$ to $2012$.}}
\end{figure}

\begin{figure}[h]
\centering
\includegraphics[width=15cm]{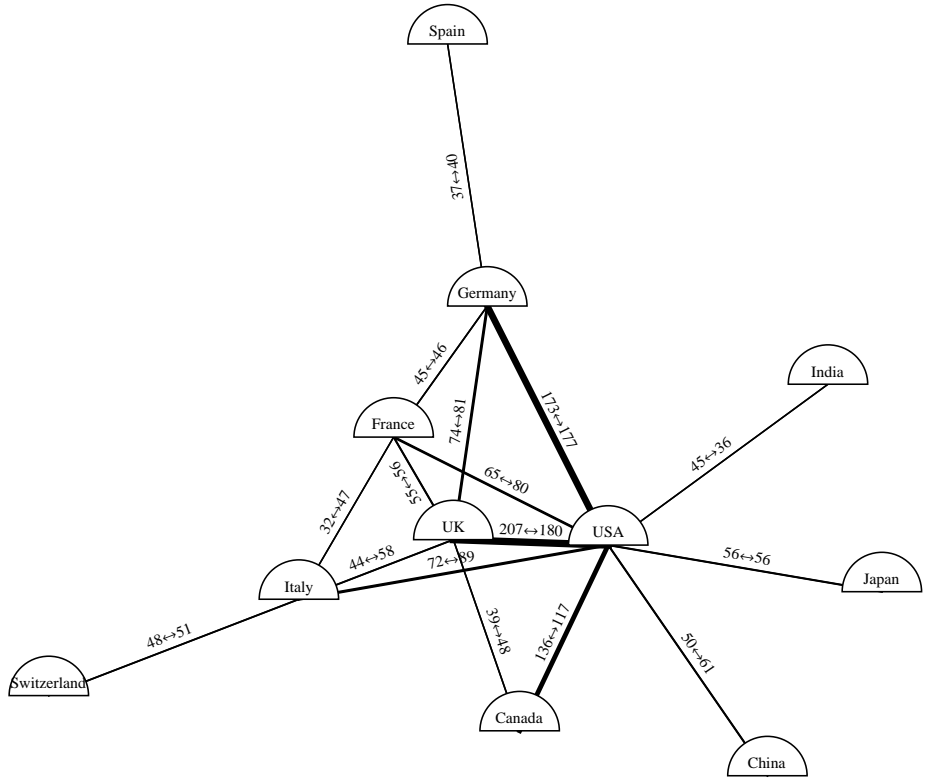}
\caption{\scriptsize{\label{fig10} Map showing the countries between which more than $30$ authors have come in and out during the years $2000$ to $2012$. Numbers are authors coming in or out a country. The arrows between them indicate the flows directions. The larger the lines, the larger the flows.}}
\end{figure}

We also evaluated authors mobility by counting the number of authors coming in and out each country from $2000$ to $2012$. By "coming in and out" we mean that an author affiliated to a country $A$ the year $Y$ and affiliated to a country $B$ the year $Y+1$ comes in $B$ but out of $A$. If $D$ is the difference between the number of authors coming in and out a country and $M$, the average number of authors coming in and out for this same country, we have calculated the ratio $D/M$ for the countries having more than $100$ authors coming in and out them from $2000$ to $2012$. This result is illustrated on figure \ref{fig9}. This ratio is negative when there is a brain drain and small when $D<<M$. The countries which attract the most of GRQC authors (i.e. that have more authors coming in than out) are Netherlands (mainly from USA and UK) and Canada (mainly from USA and UK) whereas the countries that undergo a brain drain (i.e. that have more authors coming out than in) are Italy (mainly to USA, UK, Germany, France and Spain) and India (mainly to USA).\\
Finally, we plot the graph \ref{fig10} showing the countries between which more than $30$ authors have come in and out during the years $2000$ to $2012$. The largest mobilities are between USA-Germany and USA-UK and more generally between Europe and America. Another striking feature is that there is few mobility between Europe and Asia contrarily to mobility between USA and Asia that is increasing these last years. 
\section{Conclusion and outreach} \label{s2}
The study of gr-qc electronic archive is a goldmine to measure the world effort to understand GRQC today, one hundred years after Einstein General Relativity. Let us summarise some of our main results. Europe and North America are the major places for GRQC studies. Asia and South America are rising and some seeds are sprouting in South Africa and Iran. The main journals that publish GRQC papers are PRD follows by CQG that could be overtaken in some few years by JCAP. $16$ journals publish more than $70\%$ of GRQC papers.\\
Even if a majority of papers is written by a single author, more and more papers are written by several authors and with an increasing number of international collaborations. Among all the countries, Canada is the one where the density of papers by affiliated author is the largest, Russia is the country where the largest number of papers is written by a single affiliated author whereas China is the country where an affiliated author collaborate to the most of papers each year and where papers are the most often written with authors affiliated to the same country (i.e. China). In average authors begin to find a permanent position after $6$ to $8$ years following their first publication on the archives. In average, for $30\%$ of them, this permanent position will be outside the country where they were initially affiliated $8$ years before.\\
The main collaborations between countries and main flows of authors are both between USA-UK and USA-Germany. This is to be compared to one of the classification of \cite{Eur10} that indicates that for European and US researchers, whatever the field of research, the three first countries considered in which it is most attractive to do research are USA, UK and Germany! A brain drain is observed in Italy (that was by the way number $10$ in the above-mentioned classification). Even if authors affiliated to the USA only collaborate internationally for $1/3$ of their papers, the number of their papers they deposited on the archive is so large that it is enough to make the USA a major partner for any country in the world. From this viewpoint, the number is strength.\\
We conclude by underlying that all the above results have been got by simply establishing for each research paper of an electronic archive a kind of basic id card containing the journal where it has been published (if any), its authors and the countries to which they are affiliated. If all the electronic archives where researchers upload their papers (here we consider GRQC but it might be as well in genetics, chemistry or any research field) required to fill and maintain such an id card and if all of them were freely and publicly accessible worldwide, it would be possible to deliver, at no cost, real time scientometrics for any science in any country, giving to every government a powerful tool to guide science policy.
\appendix
\section{Data strategy retrieval and data summary}\label{a0}
To retrieve the authors, with their affiliated countries, from $38291$ papers about GRQC with the journals having published them, we used the "gr-qc" papers collection of the electronic archive arXiv.org\cite{grqc}. These data can be separated in two sources.\\
The first source of data consists in the webpages of the gr-qc archive website from which we can extract, for each month of each year from $2000$ to $2012$, the reference number of the gr-qc papers, the names of their authors and the journals in which the papers have been published\footnote{For instance the webpage arxiv.org/list?year=08$\&$month=03$\&$archive=gr-qc for March 2008} (if any).\\
The second source of data are the $38291$ papers themselves under the form of LaTeX files. They can be downloaded on the AmazonS3 cloud \footnote{See arxiv.org/help/bulk$\_$data$\_$s3 and aws.amazon.com/s3/ for the download procedure}. We have analysed these LaTeX files to extract the country to which each author is affiliated for each of his/her paper. When an author indicates several affiliations corresponding to different countries, we take the first one as his/her host institution and thus affiliated country.\\
Thanks to all these data it is possible to associate to each paper of the gr-qc archive from $2000$ to $2012$, its authors, the countries to which they are affiliated and the journal in which the paper has been published (if any). This is on these data, summarized in the table below, that this study is based.
\begin{table}[H]
\begin{center}
\begin{tabular}{|c|c|c|c|c|c|}
\hline
 Year & Nb papers & Nb authors & Nb journals & $\%$ papers with & $\%$ papers with\\
 &  &  & &journal reference & countries identified \\
 \hline
 2000 & 1967 & 2129 & 98 & 76. & 98. \\
 2001 & 2055 & 2198 & 108 & 75. & 99. \\
 2002 & 2205 & 2382 & 113 & 77. & 99. \\
 2003 & 2367 & 2616 & 131 & 78. & 98. \\
 2004 & 2580 & 2809 & 136 & 77. & 98. \\
 2005 & 2683 & 3101 & 131 & 80. & 97. \\
 2006 & 2845 & 3354 & 134 & 78. & 97. \\
 2007 & 3072 & 3542 & 140 & 78. & 98. \\
 2008 & 3046 & 3777 & 161 & 78. & 99. \\
 2009 & 3383 & 4003 & 152 & 72. & 99. \\
 2010 & 3914 & 4600 & 146 & 67. & 98. \\
 2011 & 4108 & 5007 & 157 & 57. & 98. \\
 2012 & 4066 & 5181 & 137 & 47. & 99.\\
 \hline
\end{tabular}
\caption{\scriptsize{Number of papers, authors and journals identified each year on the gr-qc archive with the percentage of papers with identified journals and countries to which authors are affiliated.}}
\end{center}
\end{table}
\begin{figure}[h]
\centering
\includegraphics[width=8cm]{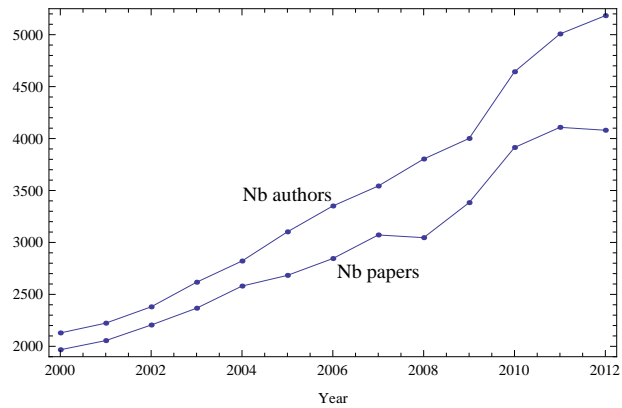}
\caption{\scriptsize{\label{fig1} Number of papers and authors from $2000$ to $2012$.}}
\end{figure}
\section{Percentages of papers edited by the $16$ journals publishing more than $1\%$ of all the papers in the gr-qc archive each year}\label{a1}
\begin{table}[H]
\begin{center}
\begin{tabular}{|r|c|c|c|c|}
\hline
& 2000 & 2004 & 2008 & 2012 \\
\hline
Physical Review D (PRD) & $30.3$ & $27.5$ & $29.5$ & $29.9$\\
Classical and Quantum Gravitation (CQG) & $13.8$ & $17.2$ & $14.6$ & $10.6$\\
General Relativity and Gravitation (GRG) & $4.7$ & $4.9$ & $3.1$ & $3.0$ \\
Physical Review Letters (PRL) & $3.7$ & $2.6$ & $2.6$ & $3.0$ \\
Physics Letters B (PLB) & $6.3$ & $4.1$ & $6.3$ & $3.8$ \\
Journal of cosmology and Astroparticle (JCAP) & $-$ & $2.9$ & $6.3$ & $6.8$ \\
Modern Physics Letter (MPL) & $2.9$ & $2.1$ & $-$ & $2.1$ \\
Journal of Hight Energy Physics (JHEP) & $-$ & $3.7$ & $5.0$ & $5.2$ \\
International Journal of Modern Physics D (IJMPD) & $3.1$ & $3.5$ & $2.6$ & $-$ \\
International Journal of Modern Physics A (IJMPA) & $-$ & $-$ & $2.1$ & $2.2$ \\
Monthly Notices of the Royal Astronomy Society (MNRAS) & $-$ & $-$ & $-$ & $3.2$ \\
European Physical Journal (EPJ) & $-$ & $-$ & $-$ & $2.2$ \\
Astrophysical Journal (APJ) & $-$ & $2.0$ & $-$ & $2.0$ \\
Journal of Mathematical Physics (JMP) & $2.4$ & $-$ & $-$ & $-$ \\
Nuclear Physics B (NPB) & $3.3$ & $-$ & $-$ & $-$ \\
Physics Letters A (PLA) & $2.5$ & $-$ & $-$ & $-$ \\
\hline
Total & $73$ & $71$ & $72$ & $74$ \\
\hline
\end{tabular}
\caption{\scriptsize{Percentages of papers edited by the $16$ journals publishing more than $1\%$ of all the papers deposited in the gr-qc archive each year}}
\label{tab2}
\end{center}
\end{table}
\section{Percentages of papers written by one author or several authors affiliated to the same or to different countries in 2012}\label{a2}
\begin{table}[H]
\begin{center}
\begin{tabular}{rcccc}
\hline
 Country &$\%$ of papers &$\%$ with several &$\%$ with several&Nb papers \\
 &with 1 author  &authors affiliated to&authors affiliated to& \\
 &  &  the same country & different countries & \\
 \hline
Argentina & 24 & 54 & 22 & 54 \\
Australia & 17 & 25 & 58 & 52 \\
Brazil & 16 & 58 & 26 & 224 \\
Canada & 25 & 25 & 50 & 260 \\
Chile & 9.9 & 45 & 45 & 71 \\
China & 13 & 65 & 22 & 228 \\
France & 24 & 24 & 53 & 228 \\
Germany & 21 & 30 & 49 & 349 \\
Greece & 24 & 26 & 50 & 58 \\
India & 27 & 48 & 25 & 241 \\
Iran & 27 & 62 & 11 & 126 \\
Israel & 39 & 27 & 33 & 51 \\
Italy & 23 & 30 & 47 & 272 \\
Japan & 19 & 44 & 37 & 228 \\
Korea & 27 & 57 & 16 & 77 \\
Mexico & 17 & 50 & 33 & 110 \\
Netherlands & 27 & 9 & 64 & 55 \\
Poland & 25 & 30 & 45 & 77 \\
Portugal & 14 & 28 & 58 & 72 \\
Russia & 38 & 28 & 35 & 138 \\
Spain & 16 & 27 & 57 & 230 \\
Switzerland & 27 & 17 & 57 & 60 \\
Taiwan & 21 & 42 & 38 & 53 \\
UK & 22 & 31 & 48 & 371 \\
USA & 29 & 39 & 32 & 831 \\
\hline
Average & 23 & 37 & 40 &\\
\hline
\end{tabular}
\caption{\scriptsize{The countries in the table are such that their affiliated authors have deposited more than $50$ papers on the gr-qc archive in $2012$ (right column). The numbers in the three first columns are, for the authors affiliated to each of these countries, the percentages of papers written in $2012$ by one author, several authors affiliated to the same country or several authors affiliated to different countries. Hence in $2012$, $29\%$ of USA papers have been written by single authors affiliated to this country, $39\%$ by several authors affiliated to this country only and $32\%$ by author(s) affiliated to this country with author(s) affiliated to other countries.}}
\label{tab3}
\end{center}
\end{table}
\bibliographystyle{unsrt}

\end{document}